\documentstyle[psfig,12pt,aaspp4]{article}

\def\zabs{$z_{\rm abs}$}
\def\zem{$z_{\rm em}$~}

\def\civ{C~{\sc iv}~}

\def\kms{$km~s^{-1}$}

\begin{document}
\title{\bf
Probing the BLR in AGNs using time variability of associated
absorption lines}
\author{R. Srianand \and S. Shankaranarayanan}
\affil{IUCAA, Post Bag 4, Ganesh Khind, Pune 411 007, 
India \\}
\authoremail{anand@iucaa.ernet.in \& shanki@iucaa.ernet.in}
\begin{abstract}

It is know that most of the clouds producing associated absorption in
the spectra of AGNs and quasars do not completely cover the background
source (continuum + broad emission line region, BLR).  We note that the
covering factor derived for the absorption is the fraction of photons
occulted by the absorbing clouds, and is not necessarily the same as
the fractional area covered. We show that the variability in
absorption lines can be produced by the changes in the covering factor
caused by the variation in the continuum and the finite light travel
time across the BLR. We discuss how such a variability can be
distinguished from the variability caused by other effects and how one
can use the variability in the covering factor to probe the BLR.

\end{abstract}


\section{Introduction}

It is known that roughly 10$\%$ of the high redshift QSOs show
associated absorption lines, ({\it i.e}  systems with absorption
redshift, \zabs, greater than or equal to the emission redshift, \zem)
super-imposed on the broad emission lines (Weymann et al., 1979; Foltz
et al., 1986; Sargent, Boksenberg \& Steidel, 1988; Barthel et al.
1991).  It is believed that these absorption systems could be due to
gas intrinsically associated with the central engine of the QSO and/or
with the ambient medium surrounding the QSO (for example interstellar
medium of the galaxies in the cluster in which QSO is also a member).
Heavy element abundance studies, using broad emission lines and
associated absorption systems in QSOs, suggest that the QSOs
are the sights of vigorous star formation activity at very early epochs
(Hamann \& Ferland, 1993; Petitjean et al., 1994; Hamann, 1997).

IUE observations suggested only 3$\%$ of the the low $z$, low
luminosity, AGNs (Seyferts) show narrow absorption lines (Ulrich,
1988).  However, HST spectra with better resolution and signal-to-noise
(S/N) show that the incidence of intrinsic absorption lines in Seyfert
1 galaxies is much higher than 3$\%$ and could be as high as 50$\%$
(Crenshaw, 1997).  Shull \& Sachs (1993) have shown the associated
\civ absorption in NGC 5548  vary over a time-scale of few days and
there is a weak anti-correlation between the continuum luminosity
and the continuum equivalent width. Such a variability in the 
associated absorption lines are also seen in few other seyfert
galaxies and QSOs (Bromage et al. 1985, walter et al. 1990,
Kolman et al. 1993, Koratkar et al. 1996, Maran et al. 1996,
Hamann et al. 1995, Weymann et al. 1997., etc.)

The variability seen in the absorption is usually assigned to
variability in the ionizing conditions in the clouds and the
time-scale of variability is used to get the density.
In this present study we propose the possibility that the variations
in the absorption line profiles over a time-scale of few days 
could be dominated by the variability in the covering factor.
We suggest continuum variability and finite light travel time across
the BLR could produce such a variability in the covering factor. 
In the following section we try to understand this using simple
illustrations.                                                                   
\section{Partial covering factor}

In this section, we formulate our approach for investigating the partial
coverage of the background emitting region by an absorbing cloud.
If the cloud is not completely covering the background source (either
continuum emitting regions [when $z_{abs}<<z_{em}$], or broad emission
line region [when $z_{abs}\simeq z_{em}$]) the observed intensity at a
wavelength $\lambda$ can be written as (Hamann et al., 1997).,
\begin{equation}
I(\lambda)~=~I_o(\lambda)(1-f_c)~+~f_c~I_o(\lambda)~exp(-\tau(\lambda)).
\end{equation}
Here $I_o(\lambda)$ is  the incident intensity,
$\tau(\lambda)$ is the optical depth and $f_c$ is the fraction of the 
background source covered by an absorbing cloud ( i.e. covering factor).
From equation (1) we can write,
\begin{equation}
\tau(\lambda)~=~-~ln~\bigg({R(\lambda)-1+f_c\over f_c}\bigg),
\end{equation}
where, $R(\lambda)$, is the measured residual intensity (i.e. $I(\lambda)/I_o(\lambda)$).
In the case of singlet line, like Ly~$\alpha$, one can estimate
the covering factor, only when the line is heavily saturated (with a flat
bottom) with some residual flux. In this case the exponential
term in the equation (1) becomes zero. However if the line is not heavily
saturated it is difficult to estimate $f_c$, as we do not know $\tau$.

In the case of doublets, however, one can calculated $f_c$ as the 
doublets ratio (the ratio of the equivalent widths of the members of 
a doublet, $W_1/W_2$) can constrain $\tau$.  The 
ratio of the oscillator strengths of the members of a doublet, ($f_1/f_2$), 
is $\sim$2 and the wavelengths ($\lambda_1,\lambda_2$) are nearly 
equal. Thus the optical depth ratio of the doublets  $(\tau_1(v)/
\tau_2(v)~=~2)$, where $v$ is the velocity with respect to the
centroid of the absorption line at which the optical depth is measured. Thus
in the case of doublets one can write,
\begin{equation}
f_c~=~{1+{R_2}^2-2R_2\over 1+R_1-2R_2},
\end{equation}
where $R_1$ and $R_2$ are the residual intensities in the first and
second lines of the doublet at a velocity $v$.  The above equation is
physical, only  when $R_2\le R_1^{0.5}$ (i.e. the covering factor
should be less than 1 and positive). In we find a situation in which
this condition is violated that will mean, (i) there is an extra
component in the first member of the doublet (due to blending effects)
and (ii) the covering factor for both the members of the doublet are
not same. The second possibility can be realized if the continuum (or
emission line)  emitting region has complicated velocity structure.
For example in the case of associated absorption lines (i.e.
$z_{abs}\simeq z_{em}$) the complicated velocity structure in the BLR
will affect our interpretation of partial coverage.

"Reverberation mapping" studies of nearby AGNs suggest that the BLR
velocities are dominated by the systemic velocity of the clouds with
high velocity components (which produce the wings of the emission line)
of the broad emission line are predominantly produced from the region
close to the continuum source and the major contribution to the low
velocity component (core of the emission line) is due to the clouds
further away from the centre. For simple illustrative purpose
we assume an ideal case in which BLR and the absorbing cloud are spherical and
they are concentric along our line of sight.
Let $r_e$ and $r_c$ are the radius of the
BLR and the absorbing cloud and  area of the background source covered
by the absorbing cloud, what we usually called covering factor, in
this case is $(r_c/r_e)^2$. However the actual covering factor at any
$v$ (with respect to the centroid of the emission line) can be derived
as follows.
\begin{equation}
I(v)~=~I_o(v,r_c>r>r_e)~+~I_o(v,<r_c)~exp[-\tau(v)]
\end{equation}
where $I_o(v,<r_c)$ and $I_o(v,r_c>r>r_e)$ are the incident 
intensities at $v$ produced by the emitting clouds within
the radius $r_c$ and in the region between $r_c$ and $r_e$ 
respectively. We define the covering factor in the 
velocity space, $\eta$ (in general case $\eta$ is nothing but
the fraction of light emitted from the continuum and BLR from
the region covered by the absorbing cloud along our line of 
sight), as,
\begin{equation}
\eta(v,r_c)~=~I_o(v,<r_c)/I_o(v,<r_e)
\end{equation}
then
\begin{equation}
I(v)~=~(1-\eta(v,r_c))~I_o(v,<r_e)+~\eta(v,r_c)~I_o(v,<r_e)~exp[-\tau(v)].
\end{equation}
The measured optical depth will be,
\begin{equation}
\tau(v)~=~-~ln~\bigg({R(v)-1+\eta(v,r_c)\over \eta(v,r_c)}\bigg)
\end{equation}
The value of $\eta(v,r_c)$ is going to depend on the radial dependent
properties of the BLR such as (i) number of line emitting clouds
per unit volume, (ii) ionization parameter (which is a dimensionless
ratio of number of ionizing photons to the number of atoms), (iii)
volume emissivity (iv) solid angle subtended by the emitting clouds on
the central continuum source etc.. In the case of associated
absorption lines the derived covering factor from the observations
$\eta(v,r_c)$ is going to be smaller than $f_c$ and the amount of
departure for a given $r_c$ is decided by the peculiar velocity of the
absorbing cloud with respect to the AGN. This effect can be understood
using a simple illustration in Figure 1. The observed emission line
(continuous curve) is decomposed into (a) emission produced within
'$r_c$' and (b) emission from $r_c<r<r_e$ in the BLR (dashed line).
For simplicity we assume the mean velocity of the
second component is much less than the first component and both
contribute equal amount to the observed intensity at $v=0$ \kms~.  If
\zabs = \zem (i.e $v_p =0$; marked as '1' in the figure)
$\eta(0,r_c)=0.50$. If the same cloud moves towards us with a velocity
$v_p=1000$ \kms~ with respect to the background source (marked as '2' in
the figure) then  $\eta(-1000,r_c)=0.75$. $\eta$ becomes 1 for any
velocity greater than 2000 \kms towards us.

Thus for a cloud covering the central region of the BLR the observed
value of $\eta$ will be high if the absorption is present in the wings
of the emission lines (with large peculiar velocity) compared to that
if the absorption is present in the core of the emission lines (with
small peculiar velocity). The case become interesting when we consider
the doublets. For example \civ doublet has a velocity separation of
$\sim$ 500 kms$^{-1}$.  Thus $\eta$'s for both the lines are going to
be different. Once again how different they can be is decided by the
peculiar velocity of the cloud with respect to the AGN and the
radial dependence of velocity distribution of the line emitting 
clouds in the BLR. The relationship
between the residual intensities measured, in the case of the doublet, is
\begin{equation}
R_2~=~1~-~\eta_2~+~\eta_2\sqrt{{(R_1-1+\eta_1)\over \eta_1}}
\end{equation}
Thus the relationship between $R_1$ and $R_2$ are not unique and one
can get different values of $R_2$ for a given value of $R_1$ by
adjusting $\eta_1$ and $\eta_2$.  Thus one needs three parameters to
fit the doublets and an unique solution is  not possible. The observed
doublet ratio and its variations  can either be a result of the
difference in $\eta$'s for the doublet lines or a genuine column
density effect. In the following subsections we discuss the
implications of these two possibilities.

\subsection {Variability in the column density}

If the ionizing continuum varies with time scale, $t_{var}$, then the
abundance of any ionization state, $i$, is given by the integrated
creation rate of the stage, $i$, extended over a time, $t_d$, that is
roughly the typical time scale for the destruction of an ion of state,
$i$, either by recombination or ionization.

\begin{itemize}
\begin{enumerate}

\item {\it when $t_d<<t_{var}$:} At any given time the ionization state
is going to be in equilibrium with the ionizing radiation. Thus one can
use simple photoionization model to predict the abundance of
different ionization states as a function of continuum variability.  In
this case we will see a clear correlation or anti-correlation between
the the continuum flux and the residual intensity (equivalent width) of
the absorption line depending upon the ionization state under
consideration.

\item {\it when  $t_d \simeq t_{var}$:} The ionic abundance follow a
history that is smoothed and delayed version of the history of the
ionizing flux. That is the residual intensity/equivalent width of the
absorption line will follow the ionizing flux with a time delay.  In
this case the ionic abundances are neither in steady state nor in
equilibrium.

\item{\it when $t_d>>t_{var}$:} The ionic abundances reaches a steady
state defined by the mean value of the ionization and recombination
rates [over the time scale $\sim t_d$], but this steady state is almost
never in equilibrium with the instantaneous value of the ionizing
flux.  In this case the equivalent width will remain constant and there
will be no clear correlation (or anti-correlation) between the
instantaneous value of the ionizing flux and the equivalent width.  One
can say there is no physical change in the absorber within the time
scale $t_{var}$.

\end{enumerate}
\end{itemize}

If the optical depth vary in response to the continuum variation over a
time scale, $t_{var}^1$, then any variation in the continuum of similar
magnitude over a time scale, $t_{var}>t_{var}^1$, should produce a
similar variation in the column density.  In the case of doublets if we
assume the variation in the covering factor are negligible then,
\begin{equation}
\dot R_1~=~{2\eta_1\over \eta_2}~e^{-\tau_1/2}~\dot R_2.
\end{equation}
As $\tau_1$ ($\sim2\tau_2$) is inversely proportional to the ionization
parameter,in the range of ionization parameter required for the
intrinsic absorption systems, one would expect to see the residual
intensity at any epoch anti-correlates with the continuum intensity and
there should be a clear correlation between $\dot R_1$ and $\dot R_2$.

\subsection {Variability in the covering factors}

If we assume the absorbing cloud covers the central source emitting the
continuum photons, then the line photon emitting clouds, within $r_c$,
respond earlier to the variation in the continuum intensity compared to
the clouds which are at a distance greater than $r_c$ from the center.
If there is a momentary increase in the ionizing continuum emission, by
$\delta(I_o)$, that will produce a momentary increase in the
$\eta(v,r_c)$. Suppose the intensity of the ionizing radiation remains
constant (at $I_o+\delta(I_o)$) for the light travel time across the
BLR  the $\eta(v,r_c)$, increases for a time equal to the light travel
time over the distance $r_c$, then falls back to its original value.
Here, we assume all the line emitting clouds are optically thick and
emit isotropically in all direction. Since the light travel time across
the BLR is 8 to12 days, any variation in the continuum intensity with a
time scale less than this will produce variation in $\eta$ that will
change the observed residual intensity in the bottom of the absorption
line.  Even if the optical depth, $\tau$, of the absorption line
remains constant (or the variation in $\tau$ is very small) one would
expect a variation in residual intensity due variation in
$\eta$ and there will  be a clear correlation between $\eta$ and residual
intensity (and hence the equivalent width). Though the variations in
the residual intensity are driven by the variations in the continuum
luminosity there will be no correlation between the continuum
luminosity and the measured residual intensity. Thus if we find the
residual intensity change when the continuum vary over a time-scale,
$t_{var}$ (which is less than the light travel time $t_c$ across
$r_c$), then any variation in the continuum of similar magnitude over a
time scale $t_{var}>t_c$ will not produce similar variation in the
residual intensity.

In the case of doublets the variation in the residual intensities,
$\dot R_1$ and $\dot R_2$, can be written in terms of variation in
$\eta$'s ($\dot\eta_1$ and $\dot\eta_2$) as, \begin{equation}
\dot{R_1}~=~\bigg{(}{{e^{-\tau_1}-1\over
e^{-\tau_2}-1}}\bigg{)}~{\dot\eta_1\over \dot\eta_2} ~\dot{R_2}
\end{equation} where $\tau_1~=~2\tau_2$. Here we have neglected the
variation in the optical depth. The value of the ratio inside the
bracket is positive and varies between 1 (when $\tau$ is very large)
and 2 (when $\tau$ is very close to zero). Thus if the ratio of
$\dot R_1$ to $\dot R_2$ is greater than 2 then we can conclude
that $\dot\eta_1>\dot\eta_2$. Similarly if the ratio is less than
1 then we can say $\dot\eta_1<\dot\eta_2$. We do not expect to see
any correlation between $\dot R_1$ and $\dot R_2$ in those cases
were the variability is dominated by the change in $\eta$'s as
the ratio of $\dot\eta_1$ by $\dot\eta_2$ need not be a constant.
On time-scales much larger than the light travel time across the
BLR the variability in the covering factor could be due to dynamical
changes in the BLR or due to the motion of the absorbing cloud
normal to the line of sight.

\subsection {Probing the BLR using covering factor variability}

If one finds a case in which the variability of the absorption line
is dominated by the variability in covering factor then in
principle using reverberation mapping techniques one can hope to infer,
\begin{itemize}
\begin{enumerate}
\item {} whether the observed variability in the covering factor can
naturally be produced by the light echo effects?
\item {} the location of the absorbing clouds, and
\item {} geometry and velocity structure of the BLR.
\end{enumerate}
\end{itemize}
 
It is usually believed that the emission line photons are produced by
the photoionization  due to UV radiation from a
central continuum source. It is a general procedure, in the
"reverberation mapping" studies, to express the observed line flux,
$F_l(t)$, at any instant $'t'$ as a function of variation in the continuum
flux ($F_c$) at earlier epochs ($t-\tau_d$), using the convolution
equation given below.

\begin{equation}
F_l(t)=<F_l>+\int_0^{\tau(max)} d\tau~\psi(\tau_d)[F_c(t-\tau_d)-<F_c>]
\end{equation}

Here, $<F_l>$, is the time averaged line flux and $<F_c>$ is the
continuum mean (as defined by Krolik \& Done, 1995) and is in general a
function of the delay, $\tau_d$.  $\psi$ is the response function (to be
obtained by inverting the above integral equation), which describes the
sensitivity of the line emissivity to continuum flux changes, integrated
over the surface of constant delay. 

The propagation of any change in the continuum through a spherical BLR
is illustrated in Figure 2. The circles represent the constant delay
surface with respect to the central source. The arrow gives the
direction of the observer and the parabola are constant time delay
surfaces (of time delay $\tau_1$, $\tau_2$ and $\tau_3$) as seen by an
observer. Obtaining, $\psi(\tau_d)$, using the emission line flux at any
velocity interval, is nothing but integrating the sensitivity to the
line emissivity of the clouds (having line of sight velocity in the
range of our interest) present in a parabola of constant delay $\tau_d$
for a continuum variation that took place $\tau_d$ days before.

Let us suppose the vertical dashed line in Figure 2 represent the
region covered by an absorbing cloud which is situated outside the BLR
(as in the previous illustrations we assume the cloud covers the
central continuum source and the central region of the BLR). In
principle, by looking at the residual intensity in the bottom of the
absorption lines, one can say whether the absorber is covering the
continuum source or not.  As can be seen from the figure, for small
values of $\tau$, the response function obtained using the line
emission, from the region covered by the absorbing cloud, will be
similar to that obtained using the emission from the whole BLR.
However at large delays,  the value of $\psi$, obtained for the covered
region will be much smaller.  Thus $\tau(max)$ needed, to fit the
variation in the line emission coming from the covered regions, will be
smaller than the $\tau(max)$ needed for the total emission. How
different they are will depend upon the distribution of line emitting
clouds along our line of sight and  their general radial distribution.
$\tau(max)$ will provide an upper limit on the size of the absorbing
region.

Since we know the emission line intensity in a
velocity interval covered by the absorption line, from the fit to the
emission line, we get the emission line intensity, at
velocity '$v$', produced by the clouds in the BLR covered by the
absorbing cloud using the calculated value of $\eta$ at that epoch.
Thus one can get the transfer function for total emission from the BLR
and the from the region covered by the absorption line. Careful
analysis of these transfer functions will give more
constraints on the BLR models than what one hope to infer from
standard reverberation mapping studies.

\section{Summary}
In this work we describe a method of investigating the partial coverage
of the background emitting region by an absorbing cloud, which will be
very much for analysing the associated absorption seen in AGNs and
QSOs. Using a simple minded example we illustrate how the
complicated velocity structure in the BLR can alter covering factor of
an absorption line. We also illustrate in general how lines of the
doublets can have different covering factor as they cover different
velocity range in a particular region of the BLR. We suggest how
one can distinguish whether the variability seen in the absorption is
due to variability in the optical depth or due to covering factor
variability.  We discuss how one can infer about the nature of the
absorption system and the BLR  using the standard reverbaration mapping
technique and the observed variability in the covering factor.



\newpage
\begin{figure}
\centerline{\vbox{
\psfig{figure=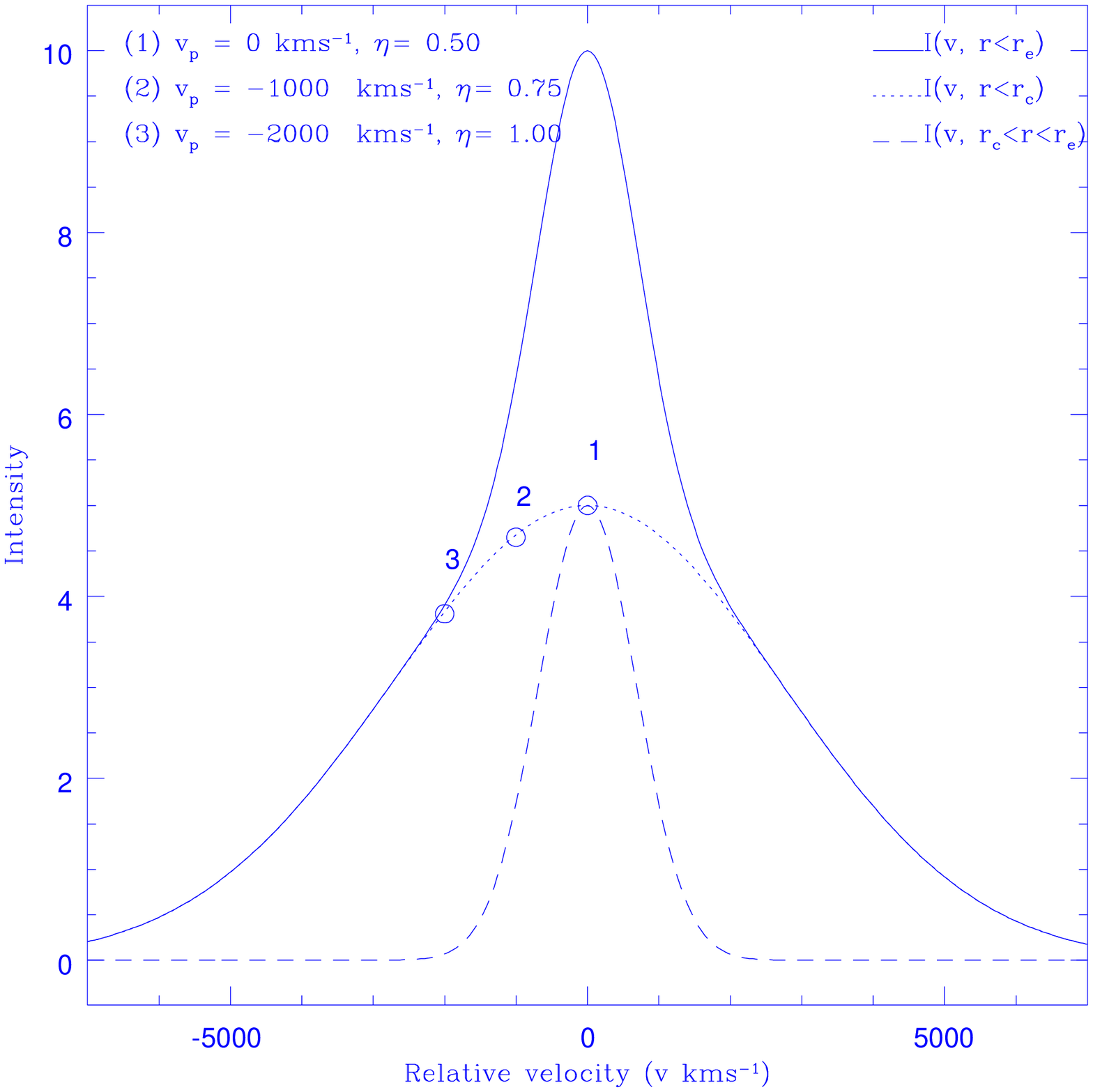,height=15cm,angle=0}
}}
\caption[]{Simple illustration of the effect of peculiar velocity
of the absorbing cloud, with respect to the AGN, on the observed
covering factor.}
\end{figure}
\begin{figure}
\centerline{\vbox{
\psfig{figure=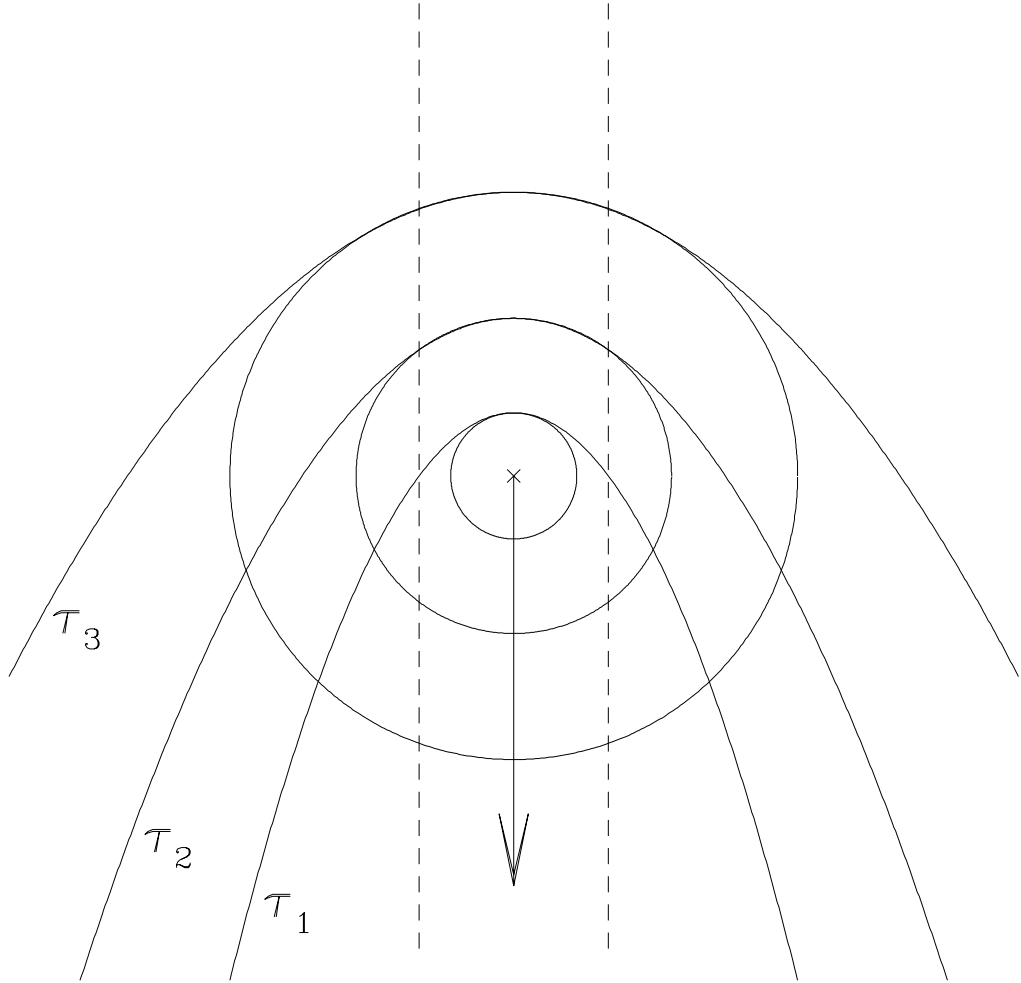,height=16cm,angle=0}
}}
\caption[]{ Schematic diagram showing the iso-delay surfaces with respect to
the central source (circles), and to an observer (parabola). The arrow 
shows the line of sight direction towards the observer and the region
between the vertical dashed lines represent the region covered by an absorbing cloud towards the observer's line of sight. }
\end{figure}
\end{document}